\documentclass[twocolumn,showpacs,preprintnumbers,amsmath,amssymb,prb,superscriptaddress]{revtex4}

\usepackage{graphicx}
\usepackage{epsf}

\begin{document}

\title{Acoustoelectric effects
in very high-mobility $p$-SiGe/Ge/SiGe heterostructure at low
temperatures in high magnetic fields}

\author{I.L.~Drichko}
\author{V.A.~Malysh}
\author{I.Yu.~Smirnov}
\affiliation{A.F. Ioffe Physical-Technical Institute of the Russian
Academy of Sciences, 194021 St.Petersburg, Russia}
\author{A.V.~Suslov}
\affiliation{National High Magnetic Field Laboratory, Tallahassee,
FL 32310, USA}
\author{O.A.~Mironov}
\affiliation{Warwick SEMINANO R\&D Centre, University of Warwick
Science Park, Coventry CV4 7EZ, UK}
\author{M.~Kummer}
\author{H.~von~K\"{a}nel}
\affiliation{Laboratorium f$\ddot{u}$r Festk\"{o}rperphysik ETH
Z\"{u}rich, CH-8093 Z\"{u}rich Switzerland}

\date{\today}

\begin{abstract}
{The contactless Surface Acoustic Wave (SAW) technique was
implemented to probe the high-frequency (ac) conductivity in a
high-mobility $p$-SiGe/Ge/SiGe structure in the integer quantum Hall
(IQHE) regime. The structure was grown by low-energy plasma-enhanced
chemical vapor deposition and comprised a two-dimensional channel
formed in a compressively strained Ge layer. It was investigated at
temperatures of 0.3 - 5.8 K and magnetic fields up to 18 T at
various SAW intensities. In the IQHE regime, in minima of the
conductivity oscillations with small filling factors, holes are
localized. The ac conductivity is of the hopping nature and can be
described within the "two-site" model. Furthermore, the dependence
of the ac conductivity on the electric field of the SAW was
determined. The manifestation of non-linear effects is interpreted
in terms of nonlinear percolation-based conductivity.}
\end{abstract}


\maketitle

\section{Introduction}
\label{Introduction} Earlier in Ref.~\onlinecite{K6016our} we
applied the acoustic method for the investigation of a high-mobility
$p$-SiGe/Ge/SiGe structure with a hole concentration of
p=6$\times$10$^{11}$cm$^{-2}$ grown by low-energy plasma-enhanced
chemical vapor deposition (LEPECVD).~\cite{lepecvd} The measurements
were performed as a function of temperature (1.6 - 4.2 K) and the
magnetic field (up to 8.4 T). We determined the general conduction
parameters of a two-dimensional hole gas in the linear regime: hole
density and mobility, effective mass, quantum and transport
relaxation times, as well as the Dingle temperature. In the
non-linear regime  we extracted the values of the energy relaxation
time and deformation potential. In our previous work, we found good
agreement between the data obtained from the contactless acoustic
technique and those from dc measurements, published in
Ref.~\onlinecite{kanel}.

The quantum Hall effect in multilayer $p$-Ge/GeSi heterostructures,
grown by the hydride method (chemical vapor deposition), was first
observed in the early 90's by the authors of
Ref.~\onlinecite{Kuznetsov}. Studies of the magnetoresistance,
carried out in a tilted magnetic field in Ref.~\onlinecite{Arapov1},
have shown that the SdH oscillation pattern of the spin subsystem
substantially depends on the degree of strain in the conducting
layers. In the case of a high compressive strain the oscillation
pattern is completely determined by the magnetic field component
normal to the sample surface.~\cite{Gorodilov,Arapov2} These authors
also pointed out the work of Ref.~\onlinecite{D'yakonov}, in which
the hole energy spectrum in the first size-quantized subband was
predicted to be significantly non-parabolic, resulting in the
dependence of the effective mass and the g-factor on the Fermi
energy, i.e., on the hole concentration.

The authors of Ref.~\onlinecite{Irisawa} studied samples with a
single $p$-Ge/GeSi quantum well, grown by molecular beam epitaxy,
and having hole concentration in the wide range of
(0.57-2.1)$\times$10$^{12}$cm$^{-2}$. They indeed observed a density
dependent effective mass. A marked dependence of the effective mass
on the concentration was also observed in Ref.~\onlinecite{Rossner}
in single $p$-Ge/GeSi quantum wells with high mobility grown by
LEPECVD. These works confirmed experimentally the prediction of
non-parabolicity of the heavy hole band in strained Ge quantum
wells.

The properties of $p$-Ge/GeSi structures were previously probed
mainly by direct current transport measurements of the
magnetoresistance or cyclotron resonance experiments. Therefore, it
was useful to investigate those structures with contactless acoustic
methods allowing calculations of the high-frequency
conductivity.~\cite{K6016our}

As the study in Ref.~\onlinecite{K6016our} was mainly focused on the
properties of a delocalized two-dimensional hole gas (2DHG), it
seemed logical to extend those measurements to identify the
low-temperature conduction mechanisms in the hole localization
regime. Moreover, the value of the g-factor is of a great interest
in these structures. In fact, the determination of this value turned
out to be complicated in Ref.~\onlinecite{K6016our}, as the spin
splitting (at odd filling numbers $\nu$=5 and 7) was barely resolved
even at $T$ = 1.6 K. Thus, the present work also addresses the
problem of determining the g-factor. For studying holes in the
localized regime and determining the g-factor, it was necessary to
carry out the measurements at lower temperatures and at higher
magnetic fields to observe oscillations corresponding to the
spin-split Landau levels with smaller filling factors and greater
oscillation amplitude.

\section{Experimental results}
\label{Experimental results}

\subsection{Object}
Here we studied the $p$-SiGe/Ge/SiGe sample (K6016) the conductivity
of which was investigated earlier by direct current~\cite{kanel} and
SAW.~\cite{K6016our} The layer structure of the sample is
illustrated in Fig.~\ref{Sample}(a).

In this system the 2D channel is formed in a compressively strained
modulation doped Ge layer. Due to the strain the threefold
degenerated valence band of Ge is split into 3 subbands, the highest
of which is occupied by heavy holes.

The entire structure was grown by LEPECVD, by making use of the
large dynamic range of growth rates offered by that
technique.~\cite{kanel} The buffer, graded at a rate of about 10$\%
/ \mu$m to a final Ge content of 70$\%$, and the 4 $\mu$m thick
constant composition layer were grown at a high rate of 5-10 nm/s
while gradually lowering the substrate temperature $T_s$ from
720$^\text{o}$C to 450$^\text{o}$C. The active layer structure,
consisting of a pure 20 nm thick Ge layer sandwiched between
cladding layers with a Ge content of about 60$\%$ and a Si cap, was
grown at a low rate of about 0.3 nm/s at $T_s$ = 450$^\text{o}$C.
Modulation doping was achieved by introducing dilute diborane pulses
into the cladding layer grown above the channel after an undoped
spacer layer of about 30 nm.

\subsection{Method} \label{Method}
\begin{figure}[h]
\centerline{
\includegraphics[width=8cm]{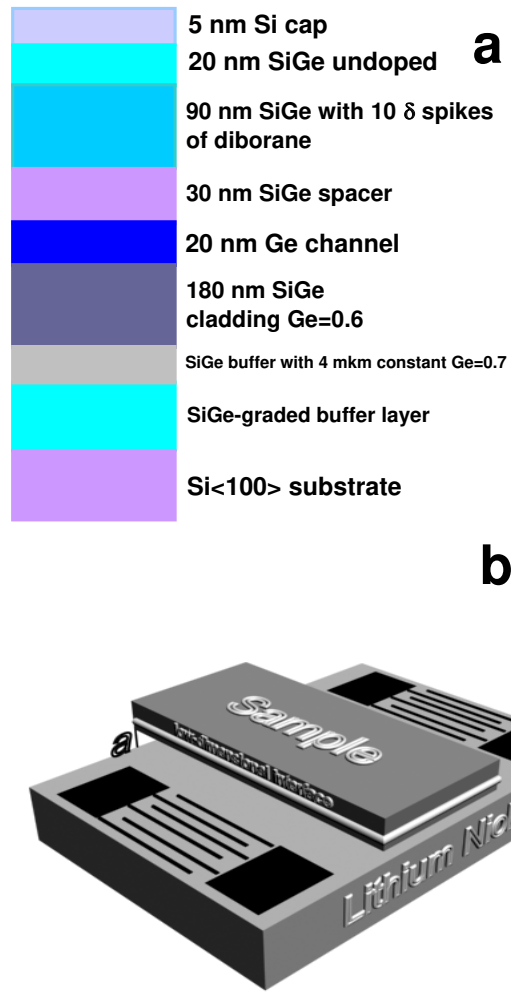}}
\caption{(a) Cross-section of the studied sample
and (b) sketch of the acoustic experiment setup.
\label{Sample}}
\end{figure}

The experimental setup is shown in Fig.~\ref{Sample}(b). A surface
acoustic wave (SAW) is excited on one side of a piezoelectric
platelet (LiNbO$_3$) by an inter-digital transducer. The SAW
propagating along the plane of the lithium niobate is accompanied by
a high-frequency electric field. This electric field penetrates into
the 2D channel located in the semiconductor structure which is
slightly pressed onto the surface of the platelet by means of
springs. The field produces electrical currents which, in turn,
cause Joule losses. As a result of the interaction of the SAW
electric field with charge carriers in the quantum well, the SAW
attenuation and its velocity change are governed by the complex
high-frequency conductivity, $\sigma^{ac}(\omega)$.

This "sandwich-like" experimental configuration enables contactless
acoustoelectric experiments on non-piezoelectric 2D systems, such as
SiGe/Ge/SiGe. The technique was first employed in
Ref.~\onlinecite{Wixforth} for GaAs/Al$_x$Ga$_{1-x}$As
heterostructures.

\subsection{Experiment} \label{Experiment}
\subsubsection{Linear regime} \label{Experiment}

\begin{figure}[ht]
\centerline{
\includegraphics[width=8cm,clip=]{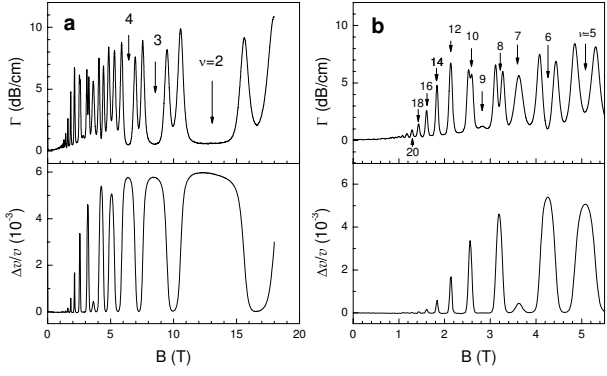}}
\caption{a) Magnetic field dependences of $\Gamma$ and
$\Delta   v/v(0)$ up to 18 T; b) $\Gamma$ and $\Delta
  v/v(0)$ in fields of up to 5.5 T. Arrows denote the filling factors. $f=30$ MHz; $T$=0.3 K.
 \label{fig:GV1}}
\end{figure}

The measurements of the SAW attenuation $\Delta \Gamma = \Gamma(B) -
\Gamma(B=0)$ and velocity change $\Delta v/v$ in the
$p$-GeSi/Ge/GeSi heterostructure, containing a single quantum well
with a hole density of p=6$\times$10$^{11}$cm$^{-2}$, were done at
frequencies of 30 and 85 MHz in magnetic fields $B$ up to 18 T and
in the temperature range of 0.3 - 5.8 K. The top panel of
Fig.~\ref{fig:GV1}a shows the magnetic field dependence of the
attenuation $\Gamma (B)$, obtained at a temperature of 0.3 K.
$\Gamma (B)$ is equal to the experimentally measured $\Delta \Gamma
(B)$, since $\Gamma(B=0) \ll \Gamma(B)$ for a zero-field
conductivity $\sigma (B=0) \approx$6$\times$10$^{-3}$ $\Omega^{-1}$.
The bottom panel of Fig.~\ref{fig:GV1}a illustrates the SAW velocity
change $\Delta v(B)/v(0) \equiv [v(B) - v(0)]/v(0)$ under the same
conditions. Analogous curves were obtained for the other
temperatures. Fig.~\ref{fig:GV1}b shows the low field portion of the
same curves corresponding to the magnetic field region from 0 to 5.5
T.
%

One can see that both the absorption coefficient and the velocity
change undergo Shubnikov-de Haas (SdH) type oscillations in the
magnetic field. The rich oscillation pattern appears at filling
factor as large as $\nu$=22.

From the measured values of $\Gamma$ and $\Delta v/v$ both the real
part $\sigma_1$ and imaginary part $\sigma_2$ of the high-frequency
(ac) conductivity $\sigma^{\text{ac}} (\omega) \equiv
\sigma_1-i\sigma_2$ could be calculated using the following
 equations taken from Ref.~\onlinecite{DrichkoPRB11}:
\begin{eqnarray}
  \label{eq:G}
&&\Gamma=8.68\frac{K^2}{2}qA     \frac{4\pi\sigma_1t(q)/\varepsilon_sv}
  {[1+4\pi\sigma_2t(q)/\varepsilon_sv]^2+[4\pi\sigma_1t(q)/\varepsilon_sv]^2}, \frac{\text{dB}}{\text{cm}}  \,   \\
&&\text{where }  A = 8b(q)(\varepsilon_1 +\varepsilon_0)
\varepsilon_0^2 \varepsilon_s
\exp [-2q(a+d)],\text{ and}  \,   \nonumber \\
\label{eq:V}
&&\frac{\Delta v}{v}=\frac{K^2}{2}A    \frac{1+4\pi\sigma_2t(q)/\varepsilon_sv}
  {[1+4\pi\sigma_2t(q)/\varepsilon_sv]^2+[4\pi\sigma_1t(q)/\varepsilon_sv]^2},
\end{eqnarray}
where $K^2$ is the electro-mechanic coupling constant for lithium
niobate (in our case, the Y-cut), $q$ and $v$ are the SAW wave
vector and velocity in LiNbO$_3$, respectively; $a$ is the gap
between the piezoelectric platelet and the sample, $d$ is the
distance between the sample surface and the 2D channel, which is
determined as described in Ref.~\onlinecite{DrichkoPRB11};
$\varepsilon_1$, $\varepsilon_0$ and $\varepsilon_s$ are the
dielectric constants of LiNbO$_3$, of
 vacuum, and of the semiconductor, respectively; $b$ and $t$ are
functions of $a$, $d$, $\varepsilon_1$, $\varepsilon_0$ and
$\varepsilon_s$.
\begin{figure}[h]
\centerline{
\includegraphics[width=8cm,clip=]{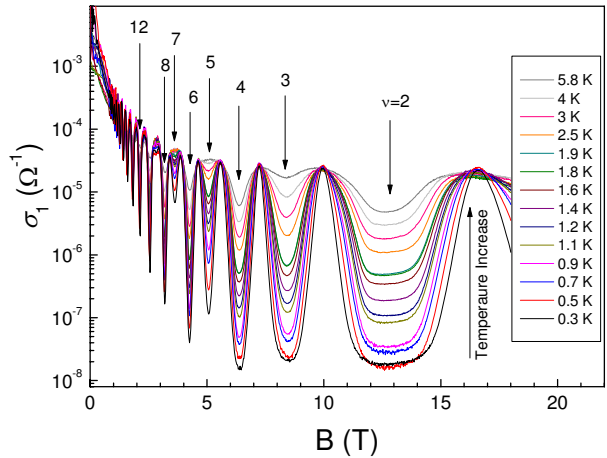}}
\caption{Dependence of $\sigma_{1}$ at 30 MHz on a magnetic field
 at different temperatures. The filling factors are marked by arrows.}
\label{S1}
\end{figure}

The magnetic field dependence of $\sigma_{1}$,  i.e., the real part
of the complex high-frequency conductivity, obtained from the SAW
measurements at different temperatures is shown in Fig.~\ref{S1}.
\begin{figure}[h]
\centerline{
\includegraphics[width=8cm,clip=]{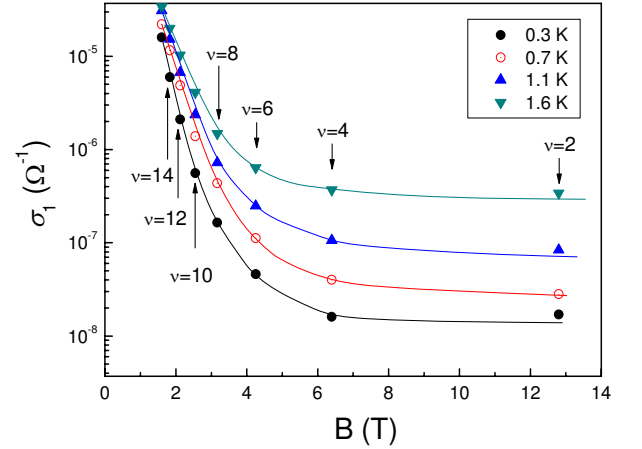}}
\caption{Magnetic field dependence of $\sigma_1$ in minima of the oscillations with
filling factors $\nu$=2, 4, 6, 8, 10, 12 and 14
at temperatures 0.3, 0.7, 1.1 and 1.6 K.
 }
\label{S1osc}
\end{figure}
\begin{figure}[h]
\centerline{
\includegraphics[width=8cm,clip=]{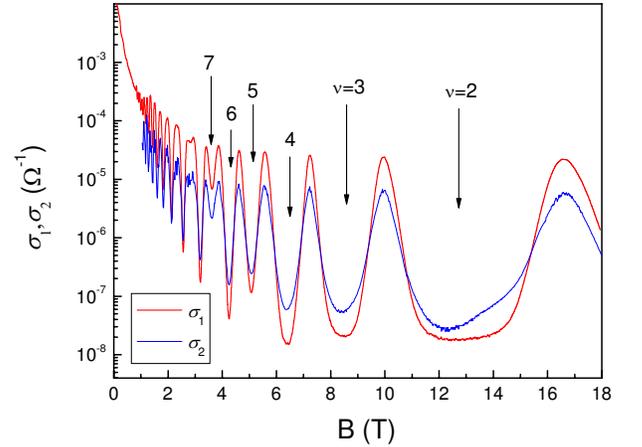}}
\caption{Magnetic field dependences of the real $\sigma_1$
and imaginary $\sigma_2$ parts of the
ac conductivity; $T$=0.3 K, $f$=30 MHz.
 }
\label{S12B}
\end{figure}

The real part of the ac conductivity in the minima of oscillations
with even filling factors at different temperatures is presented in
Fig.~\ref{S1osc}. For magnetic fields in the range of 1 T$<B<$6 T we
find that $\sigma_1 \propto 1/B^2$, while above $B>$6 T  $\sigma_1$
is virtually independent of the magnetic field.

The real $\sigma_1$ and imaginary $\sigma_2$ parts of the ac
conductivity in magnetic fields up to 18 T and at a temperature
$T$=0.3 K are displayed in Fig.~\ref{S12B}. It can be seen that in
strong fields, corresponding to small filling factors, $\sigma_1 <
\sigma_2$ in the minima of the oscillations. Yet, on the contrary,
$\sigma_1$ in its oscillation maxima tends to be greater than
$\sigma_2$. At larger filling factors the minimum of $\sigma_2$ at
first equals that of $\sigma_1$ (at $\nu =$10) and then  $\sigma_2$
becomes far smaller than $\sigma_1$ for $\nu \geq$12.
\begin{figure}[h]
\centerline{
\includegraphics[width=8cm,clip=]{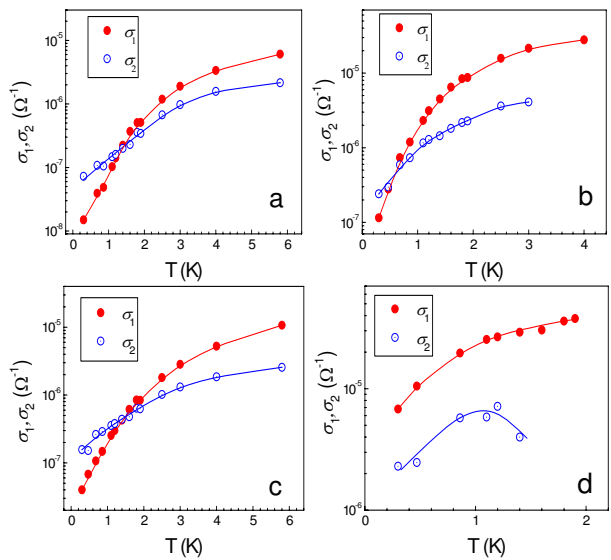}}
\caption{Temperature dependence of the real $\sigma_1$ and imaginary $\sigma_2$ parts of the
ac conductivity for different filling factors: a) $\nu$=4, b) $\nu$=5, c) $\nu$=6, d) $\nu$=7.
The lines are guides to the eye.
 }
\label{S1toS2}
\end{figure}

Figure~\ref{S1toS2} shows how the conductivities $\sigma_1$ and
$\sigma_2$ in the minima of oscillations with different filling
factors vary with the temperature.

It is seen that $\sigma_2 > \sigma_1$ for small filling factors at
low temperatures. With increasing temperature the relation between
the complex conductivity components changes, and $\sigma_1$ becomes
greater than $\sigma_2$. For filling factor $\nu$=7 $\sigma_2 <
\sigma_1$ in the whole temperature range.

The whole set of experimental dependences of the ac conductivity on
magnetic field and temperature in the oscillation minima at high
magnetic fields and low temperatures as well as the condition
$\sigma_2 > \sigma_1$ indicates a hopping nature of the
high-frequency conductivity, which can be described within a
"two-site" model.~\cite{Pollak,Efros}

In relatively small magnetic fields ($B <$ 2 T) holes are
delocalized, and $\sigma_2 \ll \sigma_1$. In this case the analysis
of the temperature damping of the SdH oscillations allows the
carrier parameters such as effective mass, Dingle temperature, and
quantum relaxation time to be evaluated. These values are close to
the ones which we determined for this sample in
Ref.~\onlinecite{K6016our}

\subsubsection{Determination of the g-factor} \label{Determination of
g-factor}

\begin{figure}[ht]
\centerline{
\includegraphics[width=8cm]{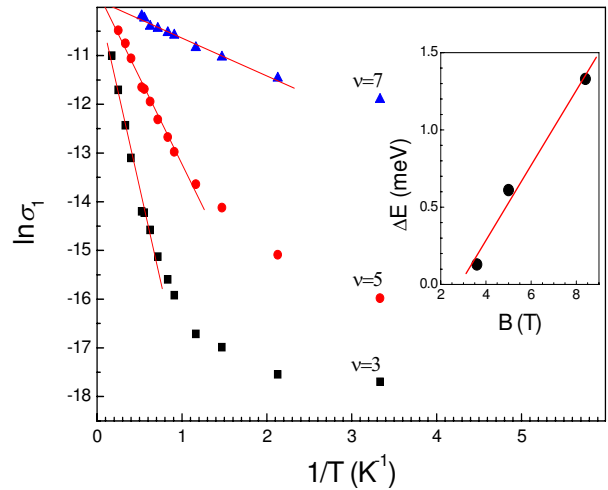}
} \caption{Arrhenius plots of $\sigma_1$ at
various odd filling factors. Inset: The activation energy in magnetic field for odd filling factors.
Lines are results of the linear fitting.
 \label{FigS1T}}
\end{figure}

To obtain the g-factor value we analyzed the temperature dependence
of the conductivity at odd filling factors $\nu$=3, 5, and 7. In the
temperature range where the Arrhenius law is valid, the real part of
the ac conductivity is described by the equation:  $\sigma_1 \propto
\exp(-\Delta E/2k_BT)$ (where $\Delta E$=$g$$\mu_B B$, $g$ is the
g-factor, $\mu_B$ is the Bohr magneton. Thus, the slope of the
linear dependence in the Arrhenius plot allows the activation energy
$\Delta E$ to be determined for each $\nu$ (Fig.~\ref{FigS1T}). The
slope of the dependences  $\Delta E (B)$, in turn, yields the value
of the g-factor. We find $g$=4.5$\pm$0.3 (see inset in
Fig.~\ref{FigS1T}).

Figure~\ref{FigS1T} demonstrates that the $\ln \sigma_1$ exhibits a
linear dependence on the inverse temperature, corresponding to an
Arrhenius law, down to temperatures of about $T \approx (0.5 \div
1)$ K. At lower temperatures the conductivity has a weaker
temperature dependence, which is usually associated with a
transition to hopping conductivity.

In the structures under study the two-dimensional channel is located
in a compressively strained Ge quantum well. It is known that in two
dimensional Ge and Si layers the compressive strain lifts the
degeneracy of the heavy and light holes of the valence band pushing
up the heavy hole energy subband. As stated in
Ref.~\onlinecite{D'yakonov}, the hole bands in the two-dimensional
objects with a structurally complicated valence band, such as thin
Ge layers, are non-parabolic. The value of the g-factor, therefore,
depends on the Fermi energy. Indeed, at the bottom of the heavy hole
band $g$=20.4,~\cite{Arapov3} while in the system with multiple
quantum wells $p$-Ge/GeSi with density
p$\approx$1$\times$10$^{11}$cm$^{-2}$ $g$ =
14$\pm$1.4,~\cite{Arapov3} to be compared with our present result of
$g$=4.5$\pm$0.3 for p=6$\times$10$^{11}$cm$^{-2}$. This is
consistent with the findings of Ref.~\onlinecite{D'yakonov}.

\subsubsection{Nonlinear regime} \label{nonlin}

The ac conductivity response to the larger power of the surface
acoustic waves, i.e., the response in the non-linear regime, was
also examined in this work.

\begin{figure}[h]
\centerline{
\includegraphics[width=8cm]{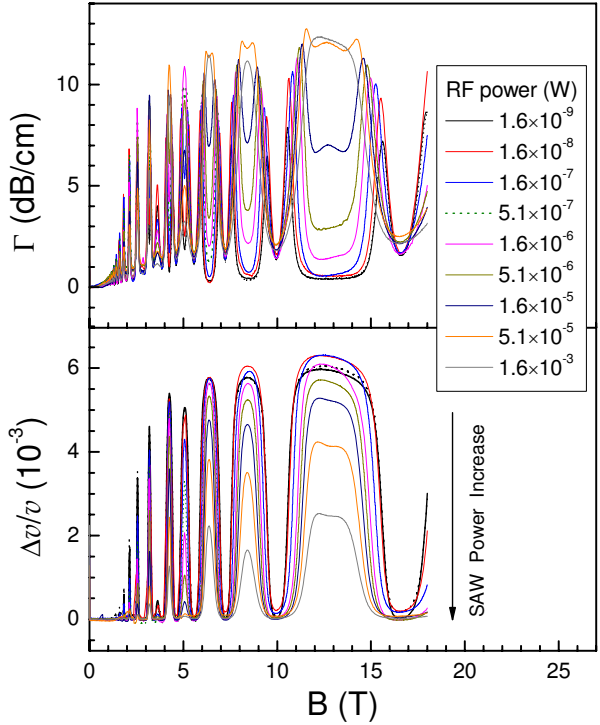}
} \caption{Dependences of $\Gamma$ and $\Delta v/v(0)$ on the magnetic field at different SAW powers;
$f$ = 30 MHz, $T$=0.3 K. \label{GamVelPower_K6016}}
\end{figure}

The absorption coefficient $\Gamma$ and velocity change $\Delta v/v$
in magnetic fields up to 18 T are displayed in
Figure~\ref{GamVelPower_K6016} at different SAW powers.

\begin{figure}[ht]
\centerline{
\includegraphics[width=8cm]{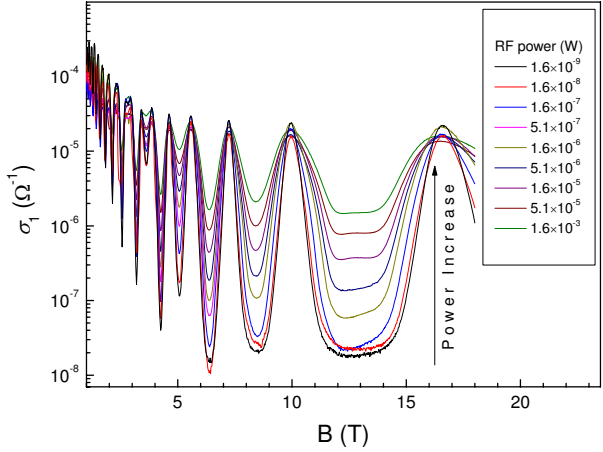}
} \caption{Dependence of $\sigma_1$ on magnetic
field at different values of RF power introduced into the sample; $T$=0.3 K. \label{S1Power}}
\end{figure}

Figure~\ref{S1Power} shows the magnetic field dependence of
$\sigma_1$ at different SAW powers, calculated using equations 1 and
2. Here it is seen that with  rising power the ac conductivity in
the oscillations minima increases while the oscillation amplitudes
decrease.

In order to identify the nature of the nonlinear effects we need to
know the electric field which accompanies the SAW. According to
Ref.~\onlinecite{DrichkoPRB11} this field can be obtained from the
following relations:
\begin{eqnarray}
  \label{eq:E}
&& |E|^2=K^2\frac{32\pi}{v}(\varepsilon_1+\varepsilon_0)     \frac{zqe^{(-2q(a+d))}} {(1+\frac{4\pi \sigma_{2}(\omega)}{\varepsilon_s
v}t)^2+(\frac{4\pi \sigma_{1}(\omega)}{\varepsilon_s v}t)^2}\frac{W}{l}  \,   , \nonumber \\
\end{eqnarray}
\begin{eqnarray}
z=[(\varepsilon_1 + \varepsilon_0)(\varepsilon_s + \varepsilon_0)-
e^{(-2qa)}(\varepsilon_1-\varepsilon_0)
(\varepsilon_s-\varepsilon_0)]^{-2}, \nonumber
\end{eqnarray}
where $W$ is the input SAW power and $l$ is the width of the sound
track.

\begin{figure}[ht]
\centerline{
\includegraphics[width=8cm]{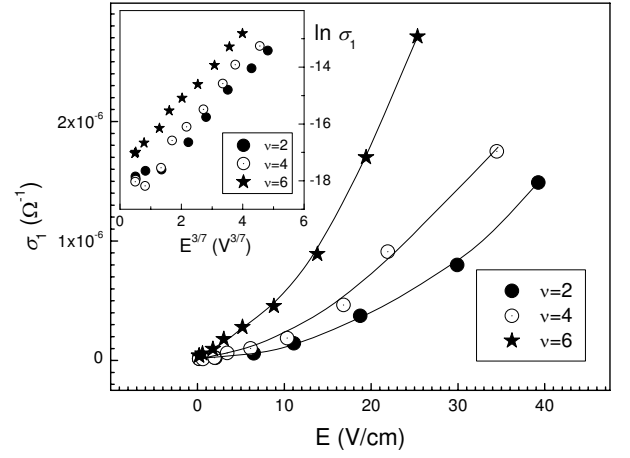}
} \caption{Real part of the ac conductivity $\sigma_{1}$
vs the SAW electric field for $\nu$=2, 4, 6; $f$=30 MHz, $T$=0.3 K.
Inset: $\ln \sigma_1$ vs $E^{3/7}$ for $\nu$=2, 4, 6. The lines are guides to the eye.\label{E3x7}}
\end{figure}

The real part of the ac conductivity $\sigma_1$ against the electric
field of the surface acoustic wave is plotted in Figure~\ref{E3x7}.

The impact of a strong constant electric field on the electrical
conductivity, caused by the activation of charge carriers to the
percolation level of the conduction band, was discussed in
Ref.~\onlinecite{Shklovskii}. The authors argued that a strong
electric field reduces the activation energy, which in turn may be
interpreted as a lowering of the percolation threshold. For the 2D
case the dependence of the conductivity on the electric field is as
follows:
\begin{equation}\label{eq4}
  \sigma_1= \sigma_1^0 \exp(\alpha E^{3/7}/k_B T),
\end{equation}

with

\begin{equation}\label{eq5}
\alpha=(Cel_{sp} V_0)^{3/7},
\end{equation}
where $\sigma_1^0$ is the conductivity in the linear regime, $C$ is
a numerical coefficient, $V_0$ is the amplitude of the fluctuations
of the random potential, and $l_{sp}$ is the characteristic spatial
scale of the potential. In our experiment the conditions
$\omega\tau_p \ll 1$ was met, where $\omega$ is the SAW frequency
and $\tau_p$ is the transport relaxation time. Therefore, the wave
(SAW) can be considered as stiffened. As a result, one may use the
theory obtained for the strong static electric fields to interpret
nonlinear effects in the ac conductivity.~\cite{Shklovskii} Although
we studied the nonlinear effects at $T$ = 0.3 K, where the
conductivity no longer has an activated character but depends weakly
on temperature (as expected for the case of "two-site" hopping), the
variation of the real conductivity $\sigma_1$ with the electric
field of the surface acoustic wave, as shown in Figure~\ref{E3x7},
is nevertheless well described by the dependence $\ln \sigma_1
\propto E^{3/7}$. One can assume that the nonlinearity mechanism is
mixed: heating the electrons in an electric field of the SAW leads
to the activated dependence of the conductivity on temperature. The
nonlinearity in this case is characterized by nonlinear conduction
at the percolation level, as in Ref.~\onlinecite{Shklovskii}.
Although this statement may require an additional proof it should be
noted that a similar behavior was observed in the non-linear hopping
conductivity in n-GaAs/AlGaAs heterostructures.~\cite{DrichkoPRB11}

\section{Conclusion} \label{Conclus}

In this paper we described the results obtained by applying
contactless acoustic techniques to the study of the high frequency
conductivity in a p-Ge/GeSi heterostructure with very high mobility.
The study was carried out at temperatures of 0.3-5.8 K in the
magnetic field of up to 18 T in both the linear and non-linear
regimes. It was shown that in the linear regime at $T <$1-2 K the
high frequency conductivity in the IQHE oscillations minima is of a
hopping nature and can be described by a "two-site" model. In the
nonlinear regime the dependence of the real part of the hopping
conductivity on the electric field of the SAW can be accurately
described by the functional form $\ln \sigma_1 \propto E^{3/7}$.

By analyzing the temperature dependences of the activated
conductivity at odd filling factors, corresponding to the spin-split
Landau levels, we were able to determine the g-factor.

\acknowledgments  This work was supported by the grant of RFBR
11-02-00223, a grant of the Presidium of the Russian Academy of
Science, the Program "Spintronika" of Branch of Physical Sciences of
RAS, grant U.M.N.I.K 16906. A portion of this work was performed at
the National High Magnetic Field Laboratory, which is supported by
NSF Cooperative Agreement No. DMR-0654118, by the State of Florida,
and by the DOE. We thank E. Palm, T. Murphy, J.H. Park, and G. Jones
for technical assistance during the experiments.

\end{document}